\documentclass[preprint,amsmath,nofootinbib,superscriptaddress,axodraw]{revtex4}
\usepackage{setspace}
\usepackage{graphicx}
\usepackage{bm}
\usepackage{epsfig}
\usepackage{color}
\usepackage{colordvi}

\newcommand{\beq}{\begin{equation}}
\newcommand{\eeq}{\end{equation}}
\newcommand{\bea}{\begin{eqnarray}}
\newcommand{\eea}{\end{eqnarray}}

\begin{document}

\title {Neutrino Mixing and the Private Higgs}

\author{Rafael A. Porto}
\affiliation{Department of Physics, University of California, Santa Barbara, CA 93106}
\author{A. Zee}
\affiliation{Department of Physics, University of California, Santa Barbara, CA 93106}
\affiliation{Kavli Institute for Theoretical Physics, University of California, Santa Barbara, CA 93106}

\begin{abstract}
We extend the Private Higgs scenario to account for charged lepton and neutrino masses by introducing a private Higgs field for each lepton family. We analyze a model in which neutrino Majorana masses are radiatively induced. Contrary to previous models the neutrino mass scale $\sim 10^{-2}$ eV is naturally obtained without excessive fine tunings. The model requires the introduction of $SU(2)$ singlet scalar fields, two charged and one neutral, $h^+$, $k^{++}$, and  $\varphi$. We discuss a possible realization giving a normal neutrino mass hierarchy. Our model fits data and predicts a non-zero $V_{e3}$. 
\end{abstract}

\maketitle

\section{Introduction}

Motivated by the disparate mass scales in the Standard Model (SM) (in which a single Higgs field has Yukawa couplings to quarks varying over a factor of ${m_t \over m_u} = {y_t \over y_u} \sim 10^5$), we introduced in a recent paper a more democratic scenario in which each quark possesses its own Private Higgs (PH), $\phi_q$, which acquires a vacuum expectation value (vev) of the order of the mass of the given quark, so that the Yukawa couplings could be comparable in magnitude \cite{ph}. Our general philosophy is that instead of having a single Higgs with vastly different couplings to the fermions, it is more natural to have many Higgs fields with similar couplings but with different vevs, which could be arranged by symmetry breaking occurring in a cascade, initiated in the specific realization given in \cite{ph} by an extra set of scalars, $S_q$, the ``darkons" \cite{xgh}, blind to the SM quantum numbers. The darkons could also provide possible candidates for the dark matter, as was suggested long ago \cite{sz} and recently studied in \cite{zee2,burg,PattWilczek,saopaulo,chris}. As we shall see the darkon partner of the electron provides the most natural of such candidates. Perhaps surprisingly, due to mixing between the different PHs in our model, the SM phenomenology with the standard Higgs is recovered, plus small corrections. 

In this paper, we extend this PH scenario to the lepton sector. Some structures similar to those in the quark sector will emerge, but also some different structures because of the Majorana character of neutrino masses. Our scheme has some attractive features regarding neutrino masses and mixing. It motivates a certain class of neutrino mass matrices, among which is for instance the matrix 
(with $m_\nu$ the scale of neutrino mass)
\[ M = m_\nu \left( \begin{array}{ccc}
\frac{1}{5} &  1 & 1  \\
 1  & 5 & -5 \\
 1 & -5 & 3 \end{array}\right)\]
which leads to the mixing matrix 
\[ |V_\nu| \sim  \left( \begin{array}{ccc}
0.84 &  0.54 & 0.02  \\
  0.33 & 0.54 & 0.77 \\
 0.43 & 0.65 & 0.63 \end{array}\right)\]
and a mass ratio $r=\frac{\Delta m^2_{\rm atm}}{\Delta m^2_{\rm solar}} \sim 0.03$, in agreement with experimental data.

A burning issue in neutrino physics is the value of $|V_{e3}|$ for which only an upper bound of $\sim 0.2$ is now known. The French experiment DoubleChooz will reach a sensitivity in the range  $\sim 0.07 - 0.09$ while the DayaBay experiment will eventually get down to $\sim 0.05$ \cite{chooz,daya}. Within this PH scheme, a range around $\sim 0.09$ (see section IV. D below) could be readily accommodated.  

An interesting feature of our model is that we have to treat the electron family as special in order for the model to work. In our previous work \cite{ph}, we found that we had to treat the top family as special. Perhaps this is consistent with a general feeling that the top among the quarks and the electron among the leptons are set apart from the others, one by virtue of its very large mass, the other by its exceptionally small mass. 

For the reader's convenience we collect the main features and formulae of the PH \cite{ph} in Appendix A. Some notations as needed will not be repeated in the text. As is typical in particle physics model building, we have a multitude of couplings. Our general philosophy, as enunciated in \cite{ph}, is that comparable couplings are roughly of the same order of magnitude. When needed, we might tolerate a ratio of $\sim10-10^2$ but not ratios of $10^5$ as in the SM. We have thus far explored only a corner of the parameter space of this class of models, and invite the reader to explore other regions.

\section{charged leptons}

In principle it would be possible to use the PHs for the down quarks $(\phi_D)$ in the lepton sector given the fact that $m_e \sim m_d$, $m_\mu \sim m_s$ and $m_\tau \sim m_b$. We will leave that possibility open for future work. Here we will privatize the charged leptons as well and introduce a PH $\phi_l$, for each one of them in the SM.  
We suppose that at some high energy scale, the individual lepton number $L_e, L_\mu, L_\tau$ are all conserved, so that the Yukawa lepton sector becomes
\begin{equation}
{\cal L}^{l}_Y = \sum_l y^{PH}_{l} {\bar \psi^{l}_L} \phi_l l_R + \mbox{h.c.}
\label{yuk}
\end{equation}
where $\psi^{l}_L \equiv (\nu_l, l)^{T}$. 
To avoid undesirable Nambu-Goldstone modes later, we might impose invariance only under a discrete subgroup of the $U(1)$'s associated with $L_e, L_\mu,$ and $L_\tau$. The strategy we will pursue in this paper is that we suppose that at some lower energy scale $L_\mu$ and $L_\tau$ are separately broken, leaving ${\tilde L} = L_\mu - L_\tau$ and $L_e$ still conserved. Finally we allow for small soft breaking of ${\tilde L}$ and subsequently of $L_e$.
To obtain the structure in (\ref{yuk}), namely to have $\phi_l$ serves only its associated lepton, we  impose two (not three) discrete symmetries $K_\tau$ and $K_\mu$ (but not $K_e$) under which
\begin{equation}
l_R\to -l_R,  ~\phi_l \to -\phi_l,~ S_l \to -S_l 
\end{equation}
for $l=\mu, \tau$.
We note that this suffices to give the desired structure in (\ref{yuk}); for instance, the term ${\bar \psi^{e}_L} \phi_\mu e_R$ is forbidden.This is similar to the set of six discrete symmetries $K_q$ we introduced in our previous work\footnote{In principle, after lepton number is broken, we could generate Flavor Changing Neutral Currents (FCNC) as in the quark sector. However, we will show these are suppressed in our model.}. 
By not introducing the $K_e$ symmetry we single out $\phi_e$ to play a special role, just as we single out $\phi_t$ in the quark sector. As mentioned above, it is perhaps natural to suspect that the electron may be special among the leptons. In fact, as we shall see, the darkon partner of the electron provides a natural candidate for dark matter.\\ 

Within the spirit of the SM we do not introduce right handed neutrinos but will seek to generate Majorana neutrino masses by radiative effects. However, let us remark that a seesaw mechanism can be easily accommodated in our framework by including $\nu_R$. In the low energy theory we can always use the PHs to construct a dimension 5 operator to generate neutrino mass.

As explained in \cite{ph} (or see the appendix), for each quark, and now for each charged lepton, we have new parameters $\gamma_{tl}$ and $M_{\phi_l}$ to play with, and thus we could make $\langle \phi^0_l \rangle$ to take on the desired order of magnitude. This general scheme could be conveniently quantified by introducing ratios similar to (\ref{tanb}) for the leptons. One salient feature of PH scheme is that the more massive the fermion, the lighter the associated PH particle. Indeed, as we showed in \cite{ph} (see also (\ref{vev0q}) below) the mass squared of the private Higgs particle for $q \neq t$ goes roughly as the inverse of the mass of the associated fermion 
\begin{equation}
M^{2}_{\phi_q} \sim \gamma_{tq} v^{3}  {\frac{1}{m_{q}}}.
\label{inverse}
\end{equation}

In contrast, since in \cite{ph} we identified the private Higgs particle associated with the top quark with the standard Higgs, its mass is given as usual by $\sim \sqrt{2\lambda_t} v$ with $v\sim 246$ GeV.

By manipulations identical to those for the quarks we obtain similarly that 
\begin{equation}
M^{2}_{\phi_l} \sim \gamma_{tl} v^{3}  {\frac{1}{m_{l}}}.
\label{lightheavy}
\end{equation}

This ``seesaw-like" relation will play a crucial role below when we try to deduce the form of the neutrino mass matrix. In estimating the contribution of a given diagram it matters greatly which PH particle circulates in the internal loop.\\

There is a subtlety regarding how the electron acquires a mass. As explained earlier and in the Appendix, $\phi^0_l$ ($l=\mu,\tau$) acquires a vev via the dimension 4 operator (see (\ref{vev0q}))
\beq
\gamma_{tl} S_t \phi^\dagger_t \phi_l S_l,
\eeq
once the darkons pick a vev and drive electroweak symmetry breaking (EWSB) (See (\ref{ls})). In principle the same mechanism would apply for the electron. However, remember we did not introduce $K_e$. Instead, we introduce now the discrete symmetry $K^s_e$ under which 
\beq
S_e \to -S_e,
\eeq
but with $\phi_e$ left unchanged. This represents a ``dark scalar version" of $K_e$.

We will assume $K^s_e$ remains unbroken, and thus the term
\beq
\label{forb}
S_e \phi^\dagger_e \phi_t S_t,
\eeq
is now forbidden. Instead we have a dimension 3 operator
\beq
\label{ge}
{\cal M} S_t\phi_t^\dagger \phi_e.
\eeq

For convenience and for uniformity of notation, we choose to write the mass parameter as ${\cal M}=\gamma_{te} v^e_s$, where $v^e_s$ replaces the role of $\langle S_e\rangle$, which we assume is zero due to $K^s_e$. The term in (\ref{ge}) induces the vev for $\phi_e^0$, and hence in practice, for the purposes of this paper, the results are unchanged and we can proceed as in the quark sector. 

The crucial assumption that $K^s_e$ remains unbroken means that $S_e$ does not get a vev and therefore the electron darkon is stable, providing a natural candidate for dark matter. In contrast, the other darkons could decay in this model.\\

In summary, the analysis in the quark sector translates literally to the $\mu,\tau$ sector. For the electron, the scalar potential is minimized at $\langle S_e\rangle = 0$. The role of (\ref{forb}) is now played by (\ref{ge}). Therefore, in the symmetry breaking cascade mentioned earlier, $S_{l\neq e}, S_t$ and $\phi_t$ acquire vevs $v^l_s/\sqrt{2}, v^t_s/\sqrt{2}$ and $v/\sqrt{2}$ respectively. Thus the linear terms $\left(\gamma_{tl}  v^l_s v^t_s v \phi_l^0\right)$ in the Lagrangian, together with the quadratic term  $M_{\phi_l}^2 \phi_l^\dagger\phi_l$ in the extended version of (\ref{ls}), lead to an expression similar to (\ref{vev0q}) with $q \to l$. Again the PH allows us to avoid unnaturally small Yukawa couplings and hence realize the motivating idea behind our scheme. For illustrative purposes we will also take $v^l_s \sim v^t_s \sim v$ for all leptons. As  far as the LHC is concerned, the PH ${\phi_\tau}$ for the $\tau$ has a mass $M_{\phi_\tau} \sim M_{\phi_b}\sim {\cal O} (1-10)$ TeV and it could play an important role in LHC physics. We intend to analyze the LHC phenomenology of our model in more detail in a future work. In contrast, $M_{\phi_e} \sim 10^3-10^4$ TeV, so that $\phi_e$ effectively decouples from the low energy theory and we only see it indirectly in the neutrino sector as we will show in what follows.

\section{Majorana neutrino masses and the private Higgs}

In the remainder of the paper we will concentrate on neutrino masses and mixing.  In order to  accommodate Majorana neutrino masses, we will ``marry" the PH model to two models introduced long ago, namely the
$\{h\phi_1\phi_2\}$ \cite{zeehpp} and $\{hhk\}$ models \cite{zeehkk,changzee,babu,santamaria}. The general philosophy behind those models is that if one particular fermion bilinear, namely $ {\bar \psi_L} l_R$, is coupled to the Higgs field, there may be no particular reason why the other fermion bilinears, $\psi_L C \psi_L$ and $ l_R C l_R$ (where $C$ stands for the charge conjugation operator), are not also coupled to scalar fields with the appropriate quantum numbers. 

\subsection{The $h$-PH model} 

In the  $\{h\phi_1\phi_2\}$ model, a charged scalar field $h^+$ is coupled to leptons according to
\begin{equation}
f_{ab} (\psi^i_{aL}C\psi^i_{bL})\epsilon_{ij} h^+ \label{neutrm}
\end{equation}
where $(i,j)$ and $(a,b)$ are $SU(2)$ and family indices respectively and Fermi statistics dictate that the coupling $f_{ab}$ is antisymmetric in $ab$. To obtain a Majorana neutrino mass we need an overall lepton number violation $\Delta L=2$. Therefore, (\ref{neutrm}) by itself does not generate a neutrino mass term since we can in principle assign $L=-2$ to $h^+$. In the original $\{h\phi_1\phi_2\}$ model, the desired lepton number violation was generated by coupling $h^+$ to two Higgs doublets $\phi_1$ and $\phi_2$.

Now the PH model enters into action. We have plenty of Higgs doublets around and we can promptly write down the cubic interaction originally proposed for the $\{ h\phi_1\phi_2\}$ model
\begin{equation}
\label{hph}
\sum_{f'f}  {\hat M}_{ff'} h^+ \phi^i_f \phi^j_{f'} \epsilon_{ij}
\end{equation}
where the sum runs over all the fermions in the SM and ${\hat M}_{ff'}$ is, due to Bose statistics, an antisymmetric ``mass" coupling. Due to this antisymmetry, the $\{h\phi_1\phi_2\}$ model requires the presence of more than one Higgs doublet, as was noted in the original paper \cite{zeehpp}. Within the PH scheme, we have plenty of Higgs doublets around for the $h^+$ field to couple to. However, the expression in (\ref{hph}) is not invariant under $K_f$ (for $f \neq e$), unless $f=f'$ in which case the expression vanishes.

Lo and behold there is a readily available way out within our model! We have our scalar darkons so that we could promote the cubic coupling above to a dimension-4 operator. We are not quite there yet, since $h^+$ is invariant under all $K_f$, and the term 
\begin{equation}
S_f h^+ \phi^i_f \phi^j_{f'} \epsilon_{ij}
\end{equation}
seems to be ruled out. Here is where our $\phi_e$ saved the day, since there is no $K_e$ and therefore the discrete symmetries $K_{\mu}, K_{\tau}, K^s_e$ and $K_q$ allow the quartic term 
\beq
\label{hph2}
\sum_f \rho_f \left(h^+ S_f  \phi^i_f \phi^j_e \epsilon_{ij}\right)
\eeq
with $\rho_{f}$ some dimensionless coupling constants. Notice that the term $f=e$ vanishes so we do not have to worry about the piece having $S_e$.

The PHs and the darkons do not carry lepton number, and hence, had we chosen to charge $h^+$ with two units of lepton number, this term (\ref{hph2}) would violate lepton number by two units. In other words, the clash between (\ref{hph2}) and (\ref{neutrm}) generates neutrino masses at one loop as in Fig 1. We denote this model as $h$-PH.\\

Notice that, after the $\phi_f$'s pick up a vev, the terms in (\ref{hph2}) produce   $\rho_t v^2 (\phi^-_e h^+)$ (since the top PH vev dominates) and thus mix $\phi_e$ and $h$. Therefore, in principle we could have a tree level contribution from this term to $\mu \to e \nu \nu$. However, the new physics contribution is suppressed by a factor of $\rho_t \frac{v^2}{M_{h^+}^2}\frac{v^2}{M_{\phi_e}^2} \sim \frac{\rho_t m_e v}{\gamma_{te}M_{h+}^2}$ with respect to the SM decay via a $W$ boson. Since we expect $M_{h^+} > v$ the new physics contribution is negligible. 

\begin{figure}[t!]
    \centering
    \includegraphics[width=9cm]{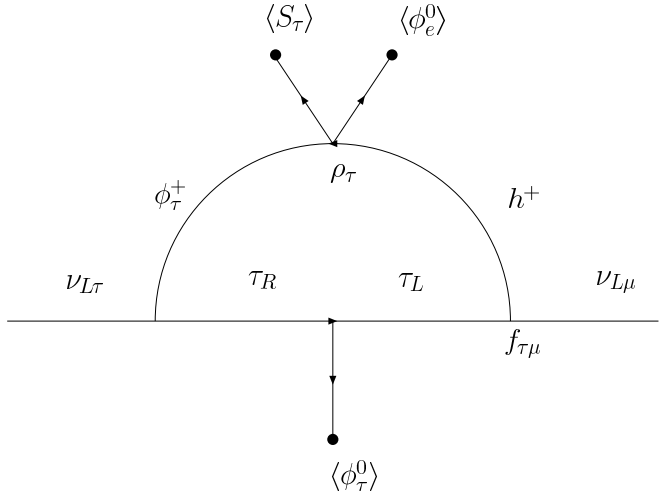}
\caption[1]{One loop contribution to the neutrino mass matrix from the $h$-PH model.}\label{nonl1}
\end{figure}

Other type of processes, such as those studied in \cite{changzee}, can be shown to be too small to be observed. For instance, The tree level contribution to neutrinoless double beta decay (see Fig (4a) in \cite{changzee}) is suppressed by $\frac{\langle \phi^0_e\rangle v}{M_{\phi_d^2}} \sim \frac{m_d m_e}{v^2} \sim 10^{-10}$. Similarly with $\mu-e$ conversion in nuclei (see Fig (4b) in \cite{changzee}).\\

Unfortunately the $h$-PH carries the seed of the simplest version of the $\{h\phi_1\phi_2\}$ model, which seems to be disfavored by experiment, since it favors the so called bi-maximal neutrino mixing matrix. In particular $V_{e2}$ comes out to be too large \cite{jarlskog, glashow, koide,haba,he}. Therefore, we will now include the scalar singlet $k^{++}$ from the $\{hhk\}$ model.

\subsection{The $\varphi k h $-PH model}

In the original $\{hhk\}$ model  \cite{zeehkk} the Lagrangian is augmented by the following terms 
\beq
{\cal L}_{hkk}= \ldots + {\tilde M} k^{++} h^-h^-  + {\tilde f}_{ab} k^{++} l^a_R C l^b_R,
\eeq
where $\tilde M$ is a mass scale. A Majorana neutrino mass is generated at two loops as shown in Fig 2.

First of all, notice that in the PH model, in contrast to the original $\{hhk\}$ model, the coupling of $k^{++}$ to the charged leptons is now constrained by the $K_{\mu,\tau}$ symmetries to be diagonal, e.g. ${\tilde f}_{ab} = {\tilde f}_a \delta_{ab}$.

\begin{figure}[t!]
    \centering
    \includegraphics[width=9cm]{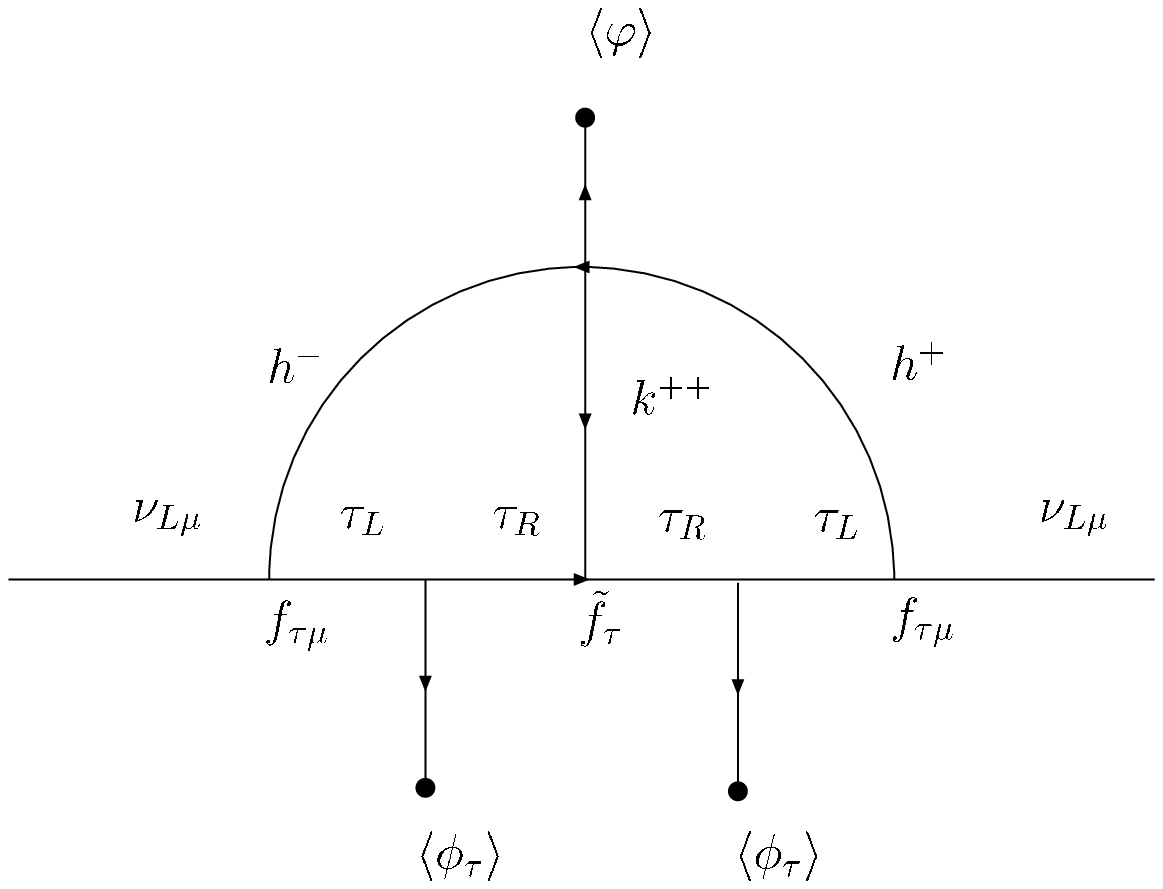}
\caption[1]{Two loops contribution to the neutrino mass matrix from the $\varphi k h$-PH model.}\label{nonl2}
\end{figure}

Secondly, in the spirit of reducing the number of dimension 3 operators, we will introduce a real scalar singlet $\varphi$ blind to the SM gauge symmetries. We denote the model as $\varphi k h$-PH, and the Lagrangian reads
\beq
{\cal L}_{\varphi khPH}= \ldots + \kappa~\varphi k^{++} h^-h^-  + \sum_a {\tilde f}_a k^{++} l^a_R C l^a_R + (\partial \varphi)^{2}-V(\varphi).
\eeq

Contributions to the neutrino mass follow at two loops once the field $\varphi$ picks a vev, e.g. $\langle \varphi \rangle = v_\varphi$. The spontaneous breaking is driven by the potential $V(\varphi)$, which also depends on other fields. As an example, let us consider a $Z_2$ discrete lepton symmetry for $\varphi$ such as $\varphi \to e^{i q_\varphi \frac{\pi}{2}} \varphi$, with $q_\varphi = 2$ (e.g. $\varphi \to -\varphi$). Therefore a natural way to introduce an instability in the $\varphi$ sector is via the terms
\beq
V(\varphi) = \ldots + \lambda_\varphi (\varphi \varphi)^2  - \lambda_{\varphi t}  \varphi \varphi \phi^\dagger_t \phi_t
- \sum_f  \lambda_{s\varphi}^f \varphi \varphi S_f S_f  + \ldots .
\eeq

For the illustrative region of parameter space we considered (recall $\langle S_f \rangle \sim v$ for all fermions other than the electron) we will naturally generate $v_\varphi \sim v$, as well as an extra scalar particle around the same scale which will also contribute to the dark sector. In what follows we will study in more detail the $\varphi k h$-PH model, in particular the neutrino mass hierarchy and mixing.

\subsection{Neutrino mass matrix and mixing}

Let us analyze in more detail the extra terms in the Lagrangian we needed to add in our $\varphi kh$-PH model, in particular regarding the symmetries of our theory. We started with three conserved lepton numbers in the charged lepton sector. 
To generate neutrino masses we needed to break lepton number by two units.  We have then a choice regarding lepton charge for the extra fields, $h,k,\varphi$, to generate neutrino masses and a phenomenologically viable mixing pattern. In principle that can also introduce non-diagonal terms in the Yukawa sector and the undesirable FCNC and decays such as $\mu \to eee$, which are strongly bounded. We will come back to this point later on.\\

With this set up, radiative corrections generate the Majorana neutrino mass term $M_{ab} \nu_{L}^{a}C\nu_{L}^{b}$ with the symmetric matrix $M_{ab}$. Within our model, there is plenty of opportunities to allow various couplings and vacuum expectation values to be complex and thus violate CP. We will not discuss CP violation here and will thus take $M_{ab}$  to be real, purely for the sake of simplicity.   

As mentioned earlier, we pursue here a natural approach of preserving $L_e$ as a symmetry to first approximation. In a sense we replace $K_e$ by $L_e$. We thus assign $L_{(e,\mu,\tau)}=0$ to $h^+$. Hence to ensure $L_e$ is conserved we are led to impose  $f_{e\mu}=0$ and $f_{e\tau}=0$. We also enforce ${\tilde f}_e=0$ by having both $\varphi$ and $k^{++}$ invariant under $L_e$. 

Notice that the symmetries we imposed imply $M_{ee}=0, M_{e\mu}=0, M_{e\tau}=0$, but $ M_{\mu\tau} \neq 0, M_{\mu\mu} \neq 0, M_{\tau\tau} \neq 0$, which leads to the normal hierarchy with $m_{1,2} \sim 0 \ll m_3$. As we will see, a very interesting pattern appears from our assumptions. 

The $f_{\mu\tau}$ coupling leaves conserved ${\tilde L} = L_\mu-L_\tau$ as alluded to earlier. However, lepton number is explicitly broken by the $\varphi k h$ sector. Since $k^{++}$ is lepton-blind, the couplings to the right handed scalar $l^a_R l^a_R$ also break ${\tilde L}$ softly. Nevertheless, notice that before symmetry breaking we can assign lepton number charges such that $L_e, L_\tau, L_\mu$ are conserved, for instance by setting ${\tilde f}_\tau=0$, as well as $L_\mu=+2$ and $-2$, for $k^{++}$ and $\varphi$ respectively. In this scenario the $\varphi k h$-PH Lagrangian preserves $L_e$ and $\tilde L$. That is enough to guarantee that FCNC are not generated before $\varphi$ picks a vev and $\tilde L$ is softly broken by the introduction of ${\tilde f}_{\tau} \neq 0$. On the other hand, Flavor Changing Charged Currents (FCCC) could be present in the $\mu-\tau$ sector like those induced from the $f_{\mu\tau}$ coupling itself. In what follows we will ignore FCCC and FCNC, we will show later on these are highly suppressed.\\

So far so good, our theory posses a $L_e$ symmetry and neutrino masses are radiatively induced with $M_{eb}=0$ for $b=e,\mu,\tau$. In principle we can make do with a neutrino mass matrix with texture $M_{eb}=0$. However, to better fit the experimental data we will relax $L_e$ and allow for a small soft breaking, but first let us get a glimpse of the order of magnitude for the entries in the neutrino mass matrix, which as we shall see are generated by different mechanisms.

In general in the original $\{h\phi_1\phi_2\}$ model, even though neutrino masses are radiatively induced, we still had to consider a very small $f_{ab}$, of the order of $10^{-5}-10^{-6}$ in (\ref{neutrm}), in order to produce values of $10^{-2}$ eV for neutrino masses. That is however not the case in the $\varphi kh$-PH model. At one loop $k^{++}$ does not enter and ${\tilde L} = L_\mu-L_\tau$ conservation implies that the only non-zero entry in the mass matrix is $M_{\tau\mu}$, from the $h$-PH piece in the Lagrangian. The diagram in Fig 1 with the $\tau$ in the loop (the same diagram with $\mu$ in the loop is down) gives
\beq
 M_{\tau\mu} \sim  {\hat f}_{\mu\tau} {\rho_\tau}\langle \phi_e \rangle \langle S_\tau \rangle\frac{m_\tau}{M_{\phi_\tau}^2} ,
\eeq
where we have introduced the reduced coupling
\beq
{\hat f}_{\mu\tau} \sim \frac{f_{\mu\tau}}{8\pi^2} \mbox{Log}\left( M_{\phi_\tau} \over  M_h \right),
\label{fhat}
\eeq
assuming $M_{\phi_\tau} > M_{h^+}$. We next use (see (\ref{vev0q}))
$M_{\phi_\tau}^2= \gamma_{t \tau} \frac{v}{\sqrt{2}} \frac{v^t_s v^\tau_s}{2\langle \phi^0_\tau \rangle } $
with $v^\tau_s \equiv \sqrt{2} \langle S_\tau \rangle$.  As explained in the Appendix we assume $v^\tau_s \sim v^t_s\sim v \sim 246$ GeV. Setting $y^{PH}_{\tau}\sim y^{PH}_{e}  \sim 1$ in
 $m_\tau \sim y^{PH}_{\tau} \langle \phi^0_\tau \rangle $ and $m_e\sim y^{PH}_{e}\langle \phi^0_e \rangle$,  we obtain
\beq
M_{\mu\tau} \sim \frac{\rho_\tau}{\gamma_{t\tau}} {\hat f}_{\mu\tau} m_e \frac{m_\tau^2}{v^2}
\label{mutau}
\eeq

A crude estimate (assuming ${\rho_\tau \sim \gamma_{t\tau}}$) sets the neutrino mass scale around
\beq
M_{\mu\tau} \sim 0.1 f_{\mu\tau}~  \mbox{eV} ,
\eeq
which naturally falls in the $10^{-2}-10^{-3} \mbox{eV}$ range with $f_{\mu\tau} \sim 10^{-1}-10^{-2}$, a much more reasonable value than the value mentioned above. The range of course depends on our various other ``reasonable" values of the other couplings in the theory. 

At this point, the neutrino mass matrix takes the form
\[ M \sim \left( \begin{array}{ccc}
0 & 0 & 0  \\
0 & 0 & x \\
0 & x & 0\end{array}\right)\]
which obviously falls short in accounting for data. The next step is to fill up the diagonal entries. Recall $L_e$ is still conserved. Thus at two loops (see Fig 2) the $k^{++}$ part of our Lagrangian will generate  non-zero values only for $M_{\mu\mu}$ and $M_{\tau\tau}$. Interestingly, these two-loop contributions could compete with $M_{\tau\mu}$. In fact, we get (assuming $M_{k^{++}}$ larger than $M_{h^+}$)
\beq
M_{\mu\mu} \sim {\tilde f}_\tau {f'}_{\mu\tau}^2 v_{\varphi} \frac{m_\tau^2}{M^2_{k^{++}}}, 
\eeq
where the reduced coupling ${f'}_{\mu\tau}$ here is  defined with the logarithmic factor Log$\left({M_{k^{++}} \over M_{h^+}}\right)$ rather than the factor in (\ref{fhat}). We will simply assume that these logarithmic factors are comparable for the sake of argument and thus  obtain (for $v_\varphi \sim v$)
\beq
\frac{M_{\mu\mu}}{M_{\tau\mu}} \sim \frac{\gamma_{t\tau}{\tilde f}_\tau {f'}_{\mu\tau}}{\rho_\tau} \frac{v^3}{m_e M^2_{k^{++}}}.
\eeq

Even though we have a $\frac{v}{m_e}$ enhancement, $M_{\mu\mu}$ is suppressed by ${\tilde f}_\tau {f'}_{\mu\tau} \sim {\tilde f}_\tau \times (10^{-3}-10^{-4})$.  Since $m_e \sim 10^{-6} v $ finally we get
\beq
\frac{M_{\mu\mu}}{M_{\tau\mu}} \sim {\tilde f}_\tau\frac{\gamma_{t\tau}}{\rho_\tau} \left(\frac{10v}{M_{k^{++}}}\right)^2,
\eeq
which can be made comparable for $M_{k^{++}} \sim  \sqrt{\tilde f_\tau} 10 v \sim $ TeV scale, but keep in mind that there are many other parameters which could change our crude estimate. 

We next turn to $M_{\tau\tau}$, generated by the diagram that produces $M_{\mu\mu}$ but with the muon in the internal line rather than the $\tau$. Thus we obtain
\beq
{M_{\tau\tau} \over M_{\mu\mu} } \sim { {\tilde f}_\mu m_\mu^2 \over {\tilde f}_\tau m_\tau^2}
\eeq 
which seems to favor a much smaller $M_{\tau\tau}$. Here we have two choices. We could either take ${\tilde f}_\mu  \sim {\tilde f}_\tau$ and hence $M_{\tau\tau}<<M_{\mu\mu}\sim M_{\mu\tau}$, or take $M_{\tau\tau} \sim M_{\mu\mu}\sim M_{\mu\tau}$, arguing that all three of these entries in the neutrino mass matrix are quantities of the same type, in contrast to quantities associated with the electron family. With various couplings around we could readily achieve this by taking ${\tilde f}_\tau \sim {\tilde f}_\mu \frac{m_\mu^2}{m_\tau^2} \sim f_{\mu\tau}$ somewhat smaller and ${\tilde f}_{\mu} \sim 1$ somewhat larger. In fact, as we discussed before, ${\tilde f}_\tau$ is responsible for the soft breaking of ${\tilde L}$ and therefore it could be naturally smaller than ${\tilde f}_\mu$. In that case, we lower the mass scale for $k^{++}$ and it could be around the corner.\\

Of course these are rough estimates which merely serve to motivate us to ``predict'' a form for the neutrino mass matrix. So far, we have filled up the lower right corner, so that
\[ M \sim \left( \begin{array}{ccc}
0 & 0 & 0  \\
0 & y & x \\
0 & x & z\end{array}\right)\]
with $x\sim y$ and either $ z \sim y $ or $z \ll y$. Now our mass matrix is closer to reproduce observations. 

Let us remark that our model has the virtue of having very little fine tuning of dimensionless constant. Indeed, the only hierarchy we stumble upon is $ f_{\mu \tau} \sim 10^{-1}-10^{-2}$, which is a small price to pay to accommodate all observed fermion masses and mixing in a compelling framework (Note in passing that $f_{\mu\tau} \sim \lambda^2-\lambda^3$, with $\lambda \sim 0.2$ from the CKM mixing.). In addition, if we take $M_{\tau\tau} \sim M_{\mu\mu}$ (that is, $y \sim z$) we would also need ${\tilde f}_\tau \sim {\tilde f}_\mu \frac{m_\mu^2}{m_\tau^2} \sim f_{\mu\tau} \sim 10^{-2} {\tilde f}_\mu$.\\ 

The final step down the road is to allow for small (soft) violations of $L_e$ to lift the zeros in our matrix up to non-zero but small values. In other words, we introduce a non-zero $f_{eb}$ such that  $ |f_{eb}| < |f_{\mu\tau}|$ for $b= \mu,\tau$. We also have to worry about FCNC, specially in the electron sector due to small violations of $L_e$, for instance in processes such as $\mu  \to eee$. However, the process $\mu \to eee$ is highly suppressed since the mass of the neutrino appears explicitly in the one-loop (see Fig. 3) generated coupling of $\phi^0_e$ to $\mu e $, giving a branching ratio of order 
\beq
\frac{A(\mu \to eee)}{A(\mu \to e \nu \nu)}  \sim {{\hat f}}_{e\mu} \frac{m_\nu m_\mu}{v^2} \frac{M^2_{W}}{M_{\phi_e}^2} \sim  {{\hat f}}_{e\mu} \frac{m_\nu m_em_\mu}{v^3} \sim f_{e\mu} 10^{-20} \sim 10^{-22}-10^{-23},
\eeq
where ${{\hat f}}_{e\mu}$ is defined similarly as in (\ref{fhat}) with $\tau \leftrightarrow \mu$, and we considered $f_{e\mu} \sim 0.1 f_{\mu\tau} \sim  10^{-2}-10^{-3}$, for which the rate is totally unobservable. Similarly for $\tau \to eee$ and other FCNC processes. For instance, the decay $\mu \to e \gamma$ is equally suppressed. The decay is described by a diagram similar to the diagram in Fig. 3 with $\phi^0_e$ replaced by its vev and a photon emitted from the $h^+$ loop.

For the case of FCCC similar considerations apply. For instance, we can generate the coupling $\phi_\mu e \bar \nu_e$, that will contribute to the dominant SM muon decay channel. The contribution is already loop suppressed ($\sim {\hat f}_{e\mu}$), and in addition we have $\frac{M_W^2}{M_{\phi_\mu}^2} \sim \frac{m_\mu}{v} \sim 4 \times 10^{-4}$. Therefore, the correction is negligible.  In principle we could also have FCCC induced by FCNC. For instance, a diagram similar to Fig. 3 generates a coupling of $\phi_\mu^0$ with $\mu e$ which in turn induces an effective coupling of the sort $\phi^+_e \mu^- \bar \nu_\mu$. Henceforth we can get contributions to $\mu^- \to \bar \nu_\mu e^- v_e$ which are experimentally constrained. However, as we have seen so far, in the PH model it follows that these effects are strongly suppressed given the high scale of the PHs, or in other words the lightness of the charged leptons involved, which renders these decays unobservable.

\begin{figure}[t!]
    \centering
    \includegraphics[width=9cm]{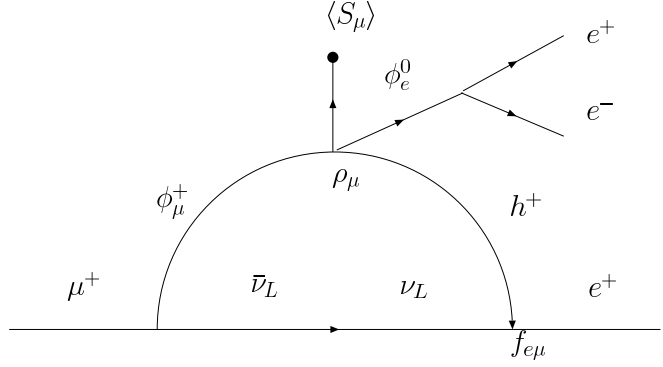}
\caption[1]{Diagram contributing to $\mu \to eee$ in the $\varphi k h$-PH model.}\label{nonl3}
\end{figure}

Adding $f_{eb}$ into the game  we end up with the following texture for the neutrino mass matrix
$$
 M \sim m_\nu \left( \begin{array}{ccc}
\delta^2 &  a \delta & b \delta  \\
 a \delta & y & x \\
 b \delta & x & z \end{array}\right)
$$
with $m_\nu \sim 10^{-2}$ eV setting the neutrino scale. Here $M_{ee}$ arises from two-loop diagrams similar to Fig 2  and it is therefore suppressed by 
$ \delta^2$ with respect to $M_{\mu\mu}$,  with $\delta \sim \frac{f_{e\tau}}{f_{\mu\tau}}$. 

On the other hand, $M_{e\mu}$ and $M_{e\tau}$ are generated in one-loop order by diagrams similar to that shown in Fig 1. Here the peculiar structure of the quartic coupling in (\ref{hph2}) gives rise to an interesting structure. The PH field $\phi_e$ associated with the electron could be either disappearing into the vacuum or circulating in the loop. In the second possibility, the largest contribution comes from $\phi_t$ (the SM Higgs) disappearing into the vacuum. We denote the two contributions respectively as $M_{e\mu}=M_{e\mu}^{(e)}+M_{e\mu}^{(t)}$ and similarly for $M_{e\tau}$. Note that for $M_{\mu\tau}$ the only possibility is for $\phi_e$ to go into the vacuum. We have 
$M_{e\mu}^{(e)} \sim \rho_\mu v {\hat f}_{e\mu} m_e m_\mu \frac{1}{M_{\phi_\mu}^2}\sim \frac{\rho_\mu}{\gamma_{t\mu} v^2} {\hat f}_{e\mu} m_e m_\mu^2$.  Similarly, $M_{e\mu}^{(t)} \sim {\rho_t \over \gamma_{te} v^2} {\hat f}_{e\mu} m_t m^2_e$ (compare with (\ref{mutau})). Thus \beq \frac{M_{e\mu}^{(t)}} {M_{e\mu}^{(e)}}\sim {\frac{\rho_t \gamma_{t\mu}}{\rho_\mu \gamma_{te} }}{\frac{m_e}{m_\mu}}{\frac{m_t}{m_\mu}}\sim 10 {\frac{\rho_t \gamma_{t\mu}}{\rho_\mu \gamma_{te} }}.\eeq 

If we take the $\rho$'s and $\gamma$'s to be comparable for lack of knowing any better, we have $M_{e\mu}\sim M_{e\mu}^{(t)}$. On the other hand, again taking the $\rho$'s and $\gamma$'s to be comparable, we have $\frac{M_{e\tau}^{(t)}}{M_{e\tau}^{(e)}}\sim {\frac{m_e}{m_\tau}}  {\frac{m_t}{m_\tau}}     \sim {\frac{1}{30}}$ and so in this case the $``(e)"$ diagram rather than the $``(t)"$ diagram dominates, giving $M_{e\tau}\sim{M_{e\tau}^{(e)}}$. We conclude that
\beq
\frac{M_{e\mu}}{M_{e\tau}} \sim  \frac    {M_{e\mu}^{(t)}}  { M_{e\tau}^{(e)}}
\sim  {\frac{m_e}{m_\tau}}        {\frac{m_t}{m_\tau}}          {\frac{ {\hat f}_{e\mu}}{ {\hat f}_{e\tau}}}                    
\sim{\frac{1}{30}}{\frac  {     {\hat f}_{e\mu}   }    {   {\hat f}_{e\tau}    }       }
\eeq
which suggests that $M_{e\mu}$ maybe somewhat smaller than $M_{e\tau}$, but since we don't know the value of the ratio 
${\frac  {     {\hat f}_{e\mu}   }    {   {\hat f}_{e\tau}    }       }
$ and of the ratios of the $\rho$'s and $\gamma$'s, not to mention ratios like $\frac{v^s_t}{v^s_e}$,  
$M_{e\mu}$ and $M_{e\tau}$ could easily be comparable.
Therefore, in our phenomenological fits below, we generally take  $M_{e\mu}$ to be somewhat smaller than or comparable to $M_{e\tau}$.

\section{Confronting experimental data}

Recall  that the neutrino mixing matrix $V$ relates the neutrino current
eigenstates (denoted by $\nu _{\alpha }$, $\alpha =e,$ $\mu ,$ $\tau ,$ to the
neutrino mass eigenstates (denoted by $\nu _{i}$, $i=1,2,3$ and with masses $m_{i}$ respectively) according to
\begin{equation}
\left( 
\begin{array}{l}
\nu _{e} \\ 
\nu _{\mu } \\ 
\nu _{\tau }
\end{array}
\right) =V\left( 
\begin{array}{l}
\nu _{1} \\ 
\nu _{2} \\ 
\nu _{3}
\end{array}
\right)  \label{vdef}
\end{equation}

We already mentioned that we are not going to study CP violation in this paper. Thus $V$ is an orthogonal matrix which diagonalizes the mass matrix, e.g. $V^T M V = \mbox{Diag} (m_1,m_2,m_3)$.\\

After nearly 50 years of extraordinary effort, experimentalists have triumphantly announced the neutrino mixing matrix \cite{splady, brlady}
\[ |V^{3\sigma}_{exp}| \sim  \left( \begin{array}{ccc}
0.77-0.86 &  0.50-0.63 & 0-0.2  \\
  0.22-0.56 & 0.44-0.73 & 0.57-0.80 \\
 0.21-0.55 & 0.40-0.71 & 0.59-0.82 \end{array}\right)\]
and the mass squared differences $\Delta m^2_{\rm solar} = m_2^2-m_1^2 \sim 8 \times 10^{-5}$ eV and
$\Delta m^2_{\rm atm} = m_3^2 - m_2^2 \sim \pm 2.4 \times 10^{-3}$ eV, hence $r=\frac{\Delta m^2_{21}}{\Delta m_{32}^2} \sim 0.03$.

We find that our model could readily fit the experimental facts.
In what follows we mention several attempts we made. As explained, our model allows for several possible choices of parameters.  We explore some of them in this section and encourage the reader to look at others.

\subsection{An attractive $M$ that almost works}

A very illuminating example follows from considering the following mass matrix
\[ M = m_\nu \left( \begin{array}{ccc}
0 &   \delta & \delta  \\
 \delta & 1 & -1/2 \\
 \delta & -1/2 & 0 \end{array}\right)\]
which complies with our theoretical bias. Notice that we take $M_{ee}$ and $M_{\tau\tau}$ to be negligible. From this simple matrix the mass ratio, $r$, comes out to be right on the money, $r = \frac{3-2\sqrt{2}}{4\sqrt{2}} + {\cal O}(\delta^2) \sim \frac{1}{33}$. In particular we also have $m_1 = 0$. Unfortunately the mixing matrix fits data only partially. The value $\delta \sim \frac{1}{5}$ seems favored and we obtain 
\[ |V_\nu| \sim  \left( \begin{array}{ccc}
0.87 &  0.48 & 0.20  \\
  0.35 & 0.27 & 0.90 \\
 0.35 & 0.85 & 0.39 \end{array}\right)\]
which has a very attractive $V_{e3} \sim 0.2$ but it is rather off in the lower right corner. Even though we should not rule it out, it seems disfavored. 

\subsection{An attractive $M$ that works}

As a second illustrative example, a very attractive fit emerges from the choice $\delta \sim \frac{1}{5}$, but $a \sim b \sim 1,~y \sim 1,~ x \sim -1$, and $z \sim \frac{3}{5}$. The latter requires a hierarchy of the sort $\frac{{\tilde f}_\mu}{{\tilde f}_\tau} \sim 10^2$. The mass matrix reads
\[ M = \frac{m_\nu}{5} \left( \begin{array}{ccc}
\frac{1}{5} &  1 & 1  \\
 1  & 5 & -5 \\
 1 & -5 & 3 \end{array}\right)\]
which leads to the mixing matrix 
\[ |V_\nu| \sim  \left( \begin{array}{ccc}
0.84 &  0.54 & 0.02  \\
  0.33 & 0.54 & 0.77 \\
 0.43 & 0.65 & 0.63 \end{array}\right)\]
and a mass ratio $r=\frac{1901+51\sqrt{1601}}{51\sqrt{1601}-1901} \sim 0.03$. In addition, we also have for the mass eigenvalues $\frac{m_3}{m_1} =  \frac{\sqrt{1601}+51}{51- \sqrt{1601}}\sim 8.3, ~\frac{m_2}{m_1} =- \frac{20}{51- \sqrt{1601}} \sim -1.8$.  Notice we have a small but non-zero $V_{e3} \sim 0.02$. Future experiments may be able to measure $V_{e3}$ down to a value $ \sim 0.01$ \cite{jl}. All the matrix elements above are within the experimental bounds \cite{splady,brlady}.

While our model does not predict the matrix $M$ given here, we might optimistically hope that a group theoretic approach or a deeper theory (perhaps string theory) would one day produce this type of matrix.

\subsection{Tri-bimaximal mixing}

A suggestive mixing matrix is the so called Tri-bimaximal matrix \cite{tbm}
\[ |V_{tbm}| = \left( \begin{array}{ccc}
\sqrt{2}/\sqrt{3} & 1/\sqrt{3} & 0  \\
 1/\sqrt{6} & 1/\sqrt{3} & 1/\sqrt{2} \\
 1/\sqrt{6} & 1/\sqrt{3} & 1/\sqrt{2} \end{array}\right)\]
which can be approximately obtained in our model with a neutrino mass matrix with $\delta = \frac{1}{5}$, $a=1$, $b=1$, $y=0.9$, $x=-1.05$, $z=0.9$. That is
\[ M = m_\nu \left( \begin{array}{ccc}
0.04 &   0.2 & 0.2  \\
 0.2 & 0.9 & -1.05 \\
 0.2 & -1.05 & 0.9 \end{array}\right)\]
where again we tuned $z \sim y$. The tri-bimaximal mixing is very appealing, however, in our model the mass ratio comes out a bit too small, e.g. $r \sim 0.018$. 

 \subsection{Making the experimentalist happy: $V_{e3} \neq 0$}

 The three examples we illustrated above have different features. However our model seems to suggest a non-zero $V_{e3}$. Very nice fits emerge from the following mass matrices, $M_A$ and $M_B$.
 
\[ M_A = m_\nu \left( \begin{array}{ccc}
0.01 &   0.05 & 0.3  \\
 0.05 & 0.85 & -1.15 \\
 0.3 & -1.15 & 1 \end{array}\right)\]
where we used $\delta = 0.1$, $a=1/2$, $b=3$, $x=-1.15$, $z=1$, $y=0.85$. The mass ratio comes out to be $r \sim \frac{1}{32}$, and the mixing matrix
\[ |V_A| \sim  \left( \begin{array}{ccc}
0.83 &  0.55 & 0.09  \\
  0.46 & 0.58 & 0.67 \\
 0.32 & 0.60 & 0.73 \end{array}\right)\]

and

\[ M_B = m_\nu \left( \begin{array}{ccc}
0.02 &  -0.06 & -0.35  \\
 -0.06 & -1 & 1.3 \\
 -0.35 & 1.3 & -1.1 \end{array}\right)\]
where we used $\delta^2 = 0.02$, $a \delta = -0.06$, $b \delta = -0.35 $, $x=1.3$, $z=-1.1$, $y=-1$. The mass ratio is  $r \sim \frac{1}{30}$, and the mixing matrix
\[ |V_B| \sim  \left( \begin{array}{ccc}
0.82 &  0.57 & 0.09  \\
  0.47 & 0.56 & 0.68 \\
 0.34 & 0.60 & 0.72 \end{array}\right)\]

In both scenario we require some tuning of $M_{\tau\tau}$ to comparable values of $M_{\mu\mu}$. However we obtain a very appealing $V_{e3} \sim 0.09$ and within the observed values for $r$.

Notice that given the observed mixing matrix and mass squared differences, we could also simply apply linear algebra and ``reverse engineer" \cite{reef,param} to obtain the neutrino mass matrix in terms of one unknown parameter.

\subsection{ $M_{ee}/m_\nu  \ll 1$}

Notice that in our model we have $\frac{M_{ee}}{m_\nu} \sim \delta^2 \ll 1$ which naturally suppresses
neutrinoless double beta decay. In addition, from the algebraic relation
 \beq
\sum_i m_i V_{ei}^2 = M_{ee} ,
\eeq
we have 
\beq
V_{e3}^2 \sim -\left( \frac{m_1}{m_3} V_{e1}^2+\frac{m_2}{m_3} V_{e2}^2\right)+\frac{M_{ee}}{m_3} \label{e3}.
\eeq

Recall $r=\frac{\Delta m^2_{21}}{\Delta m_{32}^2}$. If we now define $p=\frac{m_2}{m_3},~q=\frac{m_1}{m_3}$ we have $ r=\frac{p^2-q^2}{1-p^2}$. Within the normal mass hierarchy, $m_1 < m_2 \ll m_3$, hence $q^2 < p^2 \ll 1$, also $p<0$. Moreover, we can solve for $p(q,r)$ and plug it back in (\ref{e3}) to obtain
 \beq
V_{e3}^2 \sim \frac{M_{ee}}{m_3}+\left(\sqrt{\frac{q^2+r}{1+r}}V_{e2}^2 - q V_{e1}^2\right)\label{ve3}.
\eeq

Biased by experiments we would like the ratio $r$ to stay within values $ r \sim 0.03$, therefore we re-write $V_{e3}$ as 
\beq
|V_{e3}| \sim \left( \frac{M_{ee}}{m_3} + \frac{q}{\sqrt{1.03}} \sqrt{\left(1+\frac{0.03}{q^2}\right)} V_{e2}^2 - q V_{e1}^2\right)^{1/2}\label{ve3n}.
\eeq

Let us explore the region of parameter space where we can ignore the factor of ${M_{ee} \over m_3} \sim \delta^2 \ll 1$. Those are the cases where we expect a  large contribution to $V_{e3}$ from the other terms. We can have an idea of the values we would expect by maximizing (\ref{ve3n}) using the smaller and larger experimental values for $V_{e1}$ and $V_{e2}$ respectively (see $|V^{3\sigma}_{exp}|$ \cite{splady,brlady}). By taking $V_{e1} \sim 0.77$ and $V_{e2} \sim 0.63$ we notice that for values $0.05 < q < 0.15$  we have $0.05 < |V_{e3}| < 0.2$. Whereas on the other extreme end, with $V_{e1} \sim 0.86$ and $V_{e2} \sim 0.50$ we have $0.09 < |V_{e3}| < 0.2$ for values $0 <  q < 0.05$. Finally we can plot (\ref{ve3n}) for the central values $V_{e2}=0.565$ and $V_{e1}=0.815$, for which we have (see plot in Fig 4.)
\beq 
|V_{e3}| \sim \sqrt{0.315 q \sqrt{1+\frac{0.03}{q^2}} - 0.664 q}.
\eeq
 
Amusingly we have $V_{e3} \to 0.23$ when $q \to 0$ and $V_{e3} \to 0$ as $q$ approaches $\sim 0.1$.
 
\begin{figure}[h!]
    \centering
    \includegraphics[width=7cm]{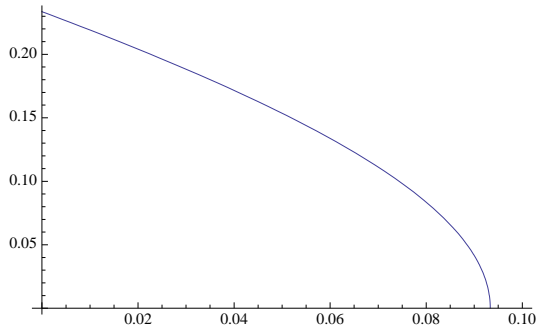}
\caption[1]{$|V_{e3}|$ as a function of $q=\frac{m_1}{m_3}$ for the central values  $V_{e2}=0.565$ and $V_{e1}=0.815$.}\label{plot}
\end{figure}

\section{Conclusions}

In this paper we extended the PH model to the lepton sector. As we did in the quark sector for the down-type quarks, we added a darkon and PH fields per lepton family such that the vevs of the latter account for the smallness of the charged lepton masses.  Similarly to the PH of the top quark, we single out the electron to play a distinctive role. To generate neutrino masses we did not include right handed neutrinos, instead we incorporated the $h^+$ and $k^{++}$ fields from the $\{h\phi_1\phi_2\}$ and  $\{hhk\}$ models, in addition to a singlet scalar $\varphi$. Majorana masses are thus induced via radiative processes. Given the plethora of new Higgses the model presents a rich structure, and motivates certain class of neutrino mass matrices. Our model fits data and suggests a non-zero $V_{e3}$. One of the main features of the PH model is the existence of the darkon fields, which provide candidates for dark matter, in particular the darkon partner of the electron which remains stable. The phenomenology of our model is plentiful and provides new signals to be explored at the LHC, such as heavy scalars charged under the SM quantum numbers and missing energy from a dark sector.  

\acknowledgments

AZ thanks Maury Goodman, John Learned, and Sandip Pakvasa for helpful discussions about the neutrino data and RAP thanks Raphael Flauger for helpful comments and discussions.
RAP is supported by the Foundational Questions Institute (fqxi.org) under grant RPFI-06-18 and the University of California and AZ in part by NSF under Grant No. 04-56556.

\appendix
\section{The Private Higgs Model in the quark sector}

Let us review the quark sector in the PH model (see \cite{ph} for details). The model is described by the Lagrangian (here ${\cal L}_{SM}$ corresponds to the SM Lagrangian with the standard Higgs $H$ set to 0) 
\begin{widetext}
\begin{eqnarray}
{\cal L} &=& {\cal L}_{SM} + \sum_q \partial_\mu S_q \partial^\mu S_q -\frac{1}{2} M_{S_q}^2 S_q^2 - \frac{\lambda_s^q}{4} S_q^4 +
\left(   (D_\mu \phi_q)^\dagger D^\mu \phi_q - \frac{1}{2} M_{\phi_q}^2 \phi_q^\dagger\phi_q - \lambda_q (\phi_q^\dagger\phi_q)^2 + g_{sq} S_q^2 \phi_q^\dagger\phi_q
 \right) \nonumber\\
&+& \sum_{q \neq q'} \left(a^s_{qq'} S_q^2 S_{q'}^2+\gamma_{qq'} S_q S_{q'} \phi_q^\dagger\phi_{q'} +  \chi_{qq'} S_{q'}^2 \phi_q^\dagger\phi_q+ a_{qq'} \phi_q^\dagger \phi_{q'} \phi_q^\dagger \phi_{q'} + b_{qq'} \phi_q^\dagger \phi_q \phi_{q' }^\dagger \phi_{q'} + c_{qq'} \phi_q^\dagger \phi_{q'} \phi_{q' }^\dagger \phi_{q}   \right)\nonumber \\ &-& \sum_q  \left( Y^{PH}_{D} {\bar Q_L} \phi_D D_q + Y^{PH}_{U} {\bar Q}_L {\tilde \phi}_U U_q\right) + \mbox{h.c.}
\label{ls}
\end{eqnarray}
\end{widetext}
where ${\tilde \phi}_q = i \sigma_2 \phi_q$, and $Y^{PH}_D, Y^{PH}_U$ are Yukawa matrices. The symmetries of the PH Lagrangian include, in addition to the SM groups, a set of six separate discrete symmetries $K_q$ 
\begin{equation}
~ U_q \to - U_q, ~ \phi_q \to -\phi_q,~ S_q \to -S_q 
\end{equation}
for $q=u,c,t$
and
\begin{equation}
D_q\to -D_q,  ~\phi_q \to -\phi_q,~ S_q \to -S_q 
\end{equation}
for $q=d,s,b$. Here $U$ and $D$ denote right handed quark fields. 

The main attractive feature of the PH is the idea of having induced fermion mass ratios, which we realize by a cascade of symmetry breaking as follows. We put the darkon fields $S_q$ in double well potentials as shown in (\ref{ls}). In contrast, we take $M_{\phi_q}^2$ positive and induce their vacuum expectation values via their interactions with the $S_q$'s. The PH partner of the top quark, namely $\phi_t$, is treated differently from the other $\phi_{q\neq t}$ PHs because of its large expectation value. (Recall that in the PH philosophy we like to have Yukawa couplings which are similar in order of magnitude while $\phi_q \sim m_q$.) We induce an effective negative mass squared term for it with the $g_{st}$ and $\chi_{tq}$ terms (See \cite{ph}.) We identify $\phi_t$ with the SM Higgs $H$. The other  $\phi_{q\neq t}$ PHs get their vev through the $\gamma_{tq}$ coupling in (\ref{ls}). Therefore, one of the main expressions behind the PH scenario is (for $q \neq t$) 
\begin{equation}
\label{vev0q}
\langle \phi^0_q \rangle = \gamma_{qt} \frac{v}{\sqrt{2}} \frac{v^t_s v^q_s}{2M_{\phi_q}^2} \sim \gamma_{qt} \frac{v}{\sqrt{2}} \frac{(v^t_s)^2}{2M_{\phi_q}^2}, 
\end{equation} 
where $v^q_s \equiv \sqrt{2} \langle S_q \rangle$ denote the vacuum expectation values of the darkon fields and we used $v^q_s \sim v^t_s$ for all $q$'s, which we took to be of the order of the electroweak symmetry breaking scale $v \sim 246$ GeV. From (\ref{vev0q}) we can naturally induce small quark masses. The smaller the mass of the quark the heavier its PH. In particular, we get the relation in (\ref{inverse}).

 We can define
\beq
\mbox{tan}\beta_q \equiv \frac{\langle \phi^0_t \rangle}{\langle \phi^0_q\rangle} = \left( \frac{2M_{\phi_q}^2}{\gamma_{qt} (v^t_s)^2}\right) \label{tanb}.
\eeq

The Yukawa couplings in the PH model can thus be compared with the SM values
\begin{equation}
y^{PH}_{q \neq t} = \mbox{tan}\beta_q ~ y^{SM}_{q\neq t}, ~~ y^{PH}_t = y^{SM}_t  \sim 1.
  \label{yph}
\end{equation}
Let $\tau_q  \sim \frac{m_t}{m_q}$, denote the desired value of $\mbox{tan}\beta_q=\frac{y^{PH}_q}{ y^{SM}_q}$, so that $\tau_q \gg 1$. Then we obtain (see \cite{ph})
\begin{equation}
\label{mph}
 M_{\phi_q} \sim  \sqrt{\frac{\tau_q}{\xi_{sqt}}} v \sim \sqrt{\gamma_{qt} \tau_q} v,
\end{equation}
where we introduced $\xi_{sqt} \equiv \frac{3g_{st}}{\lambda_t \gamma_{qt}}$. Notice that for the case of leptons we we have only $S_\tau,S_\mu$ picking up vevs and therefore the factor of $6/2=3$ in $\xi_{sqt}$, is replaced by $2/2=1$ in $\xi_{slt}$ for $l=\mu,\tau$.
As an example, for the up and down quarks we would get a PH around the $10^3$ TeV scale, whereas we can have a $\phi_b$ in the few TeV range. There are important phenomenological consequences in the PH model. The main phenomenological applications will come from the PH associated with the $b$ quark. The other Higgses will be heavy enough to `decouple' in the `low energy' regime we will explore at LHC with the exception perhaps of the PH partner of the $\tau$ lepton.

\end{document}